\documentclass[draft]{agujournal2019}
\usepackage{url} %this package should fix any errors with URLs in refs.
\usepackage{lineno}
\usepackage{soul}
\draftfalse

\journalname{Geophysical Research Letters}

\begin{document}

%% ------------------------------------------------------------------------ %%
%  Title
%
% (A title should be specific, informative, and brief. Use
% abbreviations only if they are defined in the abstract. Titles that
% start with general keywords then specific terms are optimized in
% searches)
%
%% ------------------------------------------------------------------------ %%

% Example: \title{This is a test title}

\title{Probing slow earthquakes with deep learning}

%% ------------------------------------------------------------------------ %%
%
%  AUTHORS AND AFFILIATIONS
%
%% ------------------------------------------------------------------------ %%

% Authors are individuals who have significantly contributed to the
% research and preparation of the article. Group authors are allowed, if
% each author in the group is separately identified in an appendix.)

% List authors by first name or initial followed by last name and
% separated by commas. Use \affil{} to number affiliations, and
% \thanks{} for author notes.
% Additional author notes should be indicated with \thanks{} (for
% example, for current addresses).

% Example: \authors{A. B. Author\affil{1}\thanks{Current address, Antartica}, B. C. Author\affil{2,3}, and D. E.
% Author\affil{3,4}\thanks{Also funded by Monsanto.}}

\authors{Bertrand Rouet-Leduc\affil{1}, Claudia Hulbert\affil{1,2}, Ian W. McBrearty\affil{1,3}, Paul A. Johnson\affil{1}}

% \affiliation{1}{First Affiliation}
% \affiliation{2}{Second Affiliation}
% \affiliation{3}{Third Affiliation}
% \affiliation{4}{Fourth Affiliation}

\affiliation{1}{Los Alamos National Laboratory, Geophysics Group, Los Alamos, New Mexico, USA}
\affiliation{2}{Laboratoire de G\'eologie, D\'epartement de G\'eosciences, \'Ecole Normale Sup\'erieure, PSL Research University, CNRS UMR 8538, Paris, France}
\affiliation{3}{Department of Geophysics, Stanford University, Stanford, California, USA}
%(repeat as many times as is necessary)

%% Corresponding Author:
% Corresponding author mailing address and e-mail address:

% (include name and email addresses of the corresponding author.  More
% than one corresponding author is allowed in this LaTeX file and for
% publication; but only one corresponding author is allowed in our
% editorial system.)

% Example: \correspondingauthor{First and Last Name}{email@address.edu}

\correspondingauthor{B. Rouet-Leduc}{bertrandrl@lanl.gov}

%% Keypoints, final entry on title page.

%  List up to three key points (at least one is required)
%  Key Points summarize the main points and conclusions of the article
%  Each must be 100 characters or less with no special characters or punctuation and must be complete sentences

% Example:
% \begin{keypoints}
% \item	List up to three key points (at least one is required)
% \item	Key Points summarize the main points and conclusions of the article
% \item	Each must be 100 characters or less with no special characters or punctuation and must be complete sentences
% \end{keypoints}

\begin{keypoints}
\item Deep learning models can recognize tremor on a single seismic station
\item New detections of weak tremor increase weeks to months before slow earthquakes 
\item A model trained to detect tremor in Cascadia can detect known tremor from Japan and California with no further training
\end{keypoints}

%% ------------------------------------------------------------------------ %%
%
%  ABSTRACT and PLAIN LANGUAGE SUMMARY
%
% A good Abstract will begin with a short description of the problem
% being addressed, briefly describe the new data or analyses, then
% briefly states the main conclusion(s) and how they are supported and
% uncertainties.

% The Plain Language Summary should be written for a broad audience,
% including journalists and the science-interested public, that will not have 
% a background in your field.
%
% A Plain Language Summary is required in GRL, JGR: Planets, JGR: Biogeosciences,
% JGR: Oceans, G-Cubed, Reviews of Geophysics, and JAMES.
% see http://sharingscience.agu.org/creating-plain-language-summary/)
%
%% ------------------------------------------------------------------------ %%

%% \begin{abstract} starts the second page

\begin{abstract}

Slow earthquakes may trigger failure on neighboring locked faults that are stressed enough to break, and slow slip patterns may evolve before a nearby great earthquake.
However, even in the clearest cases such as Cascadia, slow earthquakes and associated tremor have only been observed in intermittent and discrete bursts.
By training a convolutional neural network to detect known tremor on a single seismic station in Cascadia, we isolate and identify tremor and slip preceding and following known larger slow events. The deep neural network can be used for the detection of quasi-continuous tremor, providing a proxy that quantifies the slow slip rate. Furthermore, the model trained in Cascadia recognizes tremor in other subduction zones and also along the San Andreas Fault at Parkfield, suggesting a universality of waveform characteristics and source processes, as posited from experiments and theory.
\end{abstract}

\section*{Plain Language Summary}
Slow earthquakes cyclically load fault zones and have been observed preceding major earthquakes on continental faults as well as subduction zones. 
Slow earthquakes and associated tremor are common to most subduction zones, taking place down dip from the neighboring locked zone where megathrust earthquakes occur. In the clearest cases, tremor is observed in discrete bursts that are identified from multiple seismic stations.
By training a convolutional neural network to recognize known tremor on a single station in Cascadia, we detect weak tremor preceding and following known larger slow earthquakes, the detection rate of these weak tremors approximates the slow slip rate at all times, and the same model is able to recognize tremor from different tectonic environments with no further training.

%% ------------------------------------------------------------------------ %%
%
%  TEXT
%
%% ------------------------------------------------------------------------ %%

%%% Suggested section heads:
% \section{Introduction}
%
% The main text should start with an introduction. Except for short
% manuscripts (such as comments and replies), the text should be divided
% into sections, each with its own heading.

% Headings should be sentence fragments and do not begin with a
% lowercase letter or number. Examples of good headings are:

% \section{Materials and Methods}
% Here is text on Materials and Methods.
%
% \subsection{A descriptive heading about methods}
% More about Methods.
%
% \section{Data} (Or section title might be a descriptive heading about data)
%
% \section{Results} (Or section title might be a descriptive heading about the
% results)
%
% \section{Conclusions}

\section*{Introduction}

In many subduction zones where megaquakes occur, slow earthquakes take place deep on the subduction interface, perturbing the stress environment of the neighboring and shallower locked zone, and potentially influencing the occurrence of large megathrust earthquakes\cite{ITO2013,VidaleHouston}. As a result, the relation between slow earthquakes and the locked zone is a topic of intense interest. Non-volcanic tremor, the noise-like seismic signature of slow earthquakes emanating from the subduction zone, was discovered relatively recently \cite{Obara2002,Rogers2003}. % owing to its very unremarkable nature on a single station. 
Tremor is identified and located by analyzing the phase correlations of seismic signal envelopes recorded at multiple seismometers \cite{Obara2002,Obara2016}.  Tremor is frequently used as a qualitative proxy for slow slip \cite{VidaleHouston,Ide2012}, including in situations where GPS shows no displacement, presumably because deformation at the surface is too small to record  \cite{Frank2016}. 

% We rely on convolutional neural networks (CNNs), a type of deep learning algorithm, in an effort to characterize tremor from seismic noise and probe the relationship between tremor and slip (Fig. 1).  

In an effort to characterize tremor from seismic noise and probe the relationship between tremor and slip (Fig. 1), we use a method based on convolutional neural networks (CNNs), a type of deep learning algorithm. CNNs are at the core of recent dramatic advances in computer vision, natural language processing  and recommender systems \cite{LeCun2015}. The use of CNNs to recognize tremor can be viewed as a deep learning extension of template matching methods \cite{Gibbons2006, Frank2014}, where the deep learning model automatically determines which time-frequency patterns to use, effectively learning something akin to a set of templates that are more general representations of tremor than hand-crafted templates. Tasked with recognizing tremor from portions of single station seismic data, the convolutional layers learn to represent inputs as a collection of simpler characteristic features. These features are then fed to the dense layers of the CNN, that learns to classify whether a portion of seismic data contains tremor or not based on this transformed representation.

% learn to transform this data into simpler characteristic features

\section*{Results}

\subsection*{Deep learning tremor}

The CNN model is trained on seismic data from the Canadian National Seismograph Network (CNSN) \cite{CNSN}.  Figure 2A shows the area analyzed: Vancouver Island on the North American Plate and the subducting Juan de Fuca plate, with a schematic of the locked and slowly slipping portions of the downgoing slab. Our ground truth for the initial training of the neural network is a tremor catalog from the Pacific Northwest Seismic Network (PNSN). The catalog was constructed using Wech's tremor identification method (multi-station, based upon 5 minute envelope correlation) \cite{wech2008} from the southern portion of Vancouver Island between October 2009 to July 2017. We build our database from 5 minute portions of single station seismograms. For every tremor event in the PNSN catalog, we record the corresponding 5 minute single station waveform, and label it as containing tremor. For the non-tremor examples we randomly sample the seismic data on days where no tremor was identified in the PNSN catalog. This results in about 47,500 time windows of 5 minute single station waveforms labeled as containing tremor, and 47,500 windows of 5 minute single station waveforms labeled as \textit{not} containing tremor.  In order to leverage the ability of CNNs to extract information from images, instead of feeding raw waveforms to the CNN, we first compute the short-time Fourier transform (STFT) of the 5 minute portions of data (Fig. 1B). The original 5 minute waveforms containing 12'000 data points are converted into spectrograms over the 5 minute interval. These steps result in our database of 95,000 labeled `tremor' and `absence of tremor' examples (Fig. 1D,E).

% The spectral image of each 5 minute single station waveform is in turn labeled with the corresponding `tremor' or `absence of tremor' as determined by the multi-station PNSN catalog to constitute our database of 95,000 labeled examples (Fig. 1D,E). 

\begin{figure}[!ht]
\begin{center}
\includegraphics[width=14cm,trim= 0 0 0 0]{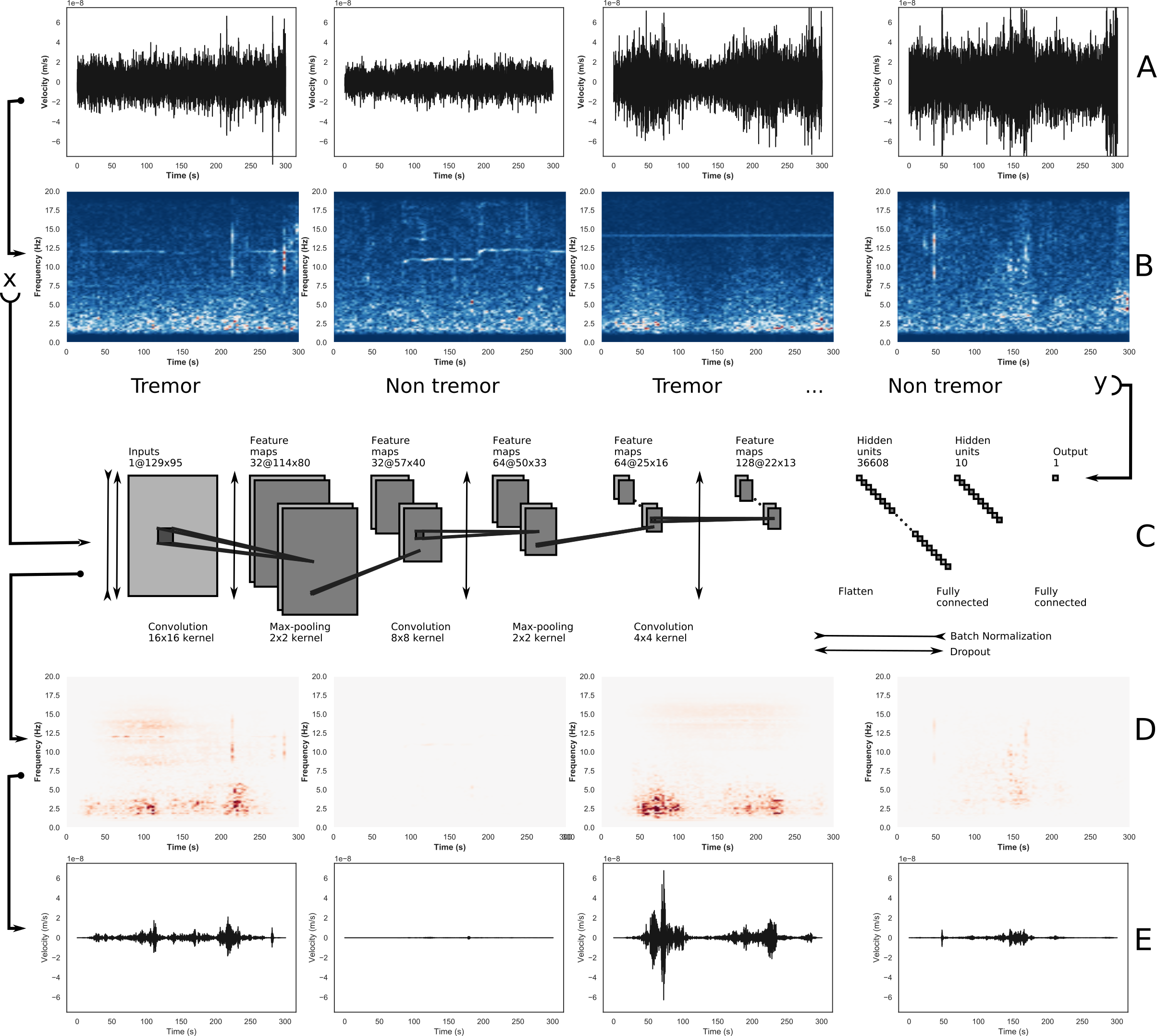}
\caption{\footnotesize{\textbf{Deep learning tremor.} \textbf{(A)} 5-minute duration, single station (NLLB) seismic signals. \textbf{(B)} Short-time Fourier transform of the waveforms that are fed as input to the convolutional neural network. \textbf{(C)} Schematic of the CNN and its architecture. The convolutional layers learn representations (features) of tremor while the last dense layers determine detection/no detection of tremor based on the presence of these features in a spectrogram. The model is trained on spectrograms labeled using the PNSN catalog of tremor from southern Vancouver Island from 2009 to 2015. \textbf{(D)} Interpretation of the CNN using Taylor decomposition\cite{Montavon2017}, showing in red which parts of the spectrograms were recognized as characteristic of tremor. \textbf{(E)} Reconstruction of the waveforms from only the portions of the spectrograms recognized as tremor according to the CNN and its interpretation.}}
\label{fig1}
\end{center}
\end{figure}

We split our database of 95'000 time (frequency) windows into two contiguous portions for training and testing: the first 80$\%$ for training and the last 20$\%$ for validation (10$\%$) and testing (10$\%$). In other words, all the examples from the end of 2009 to the end of 2015 are used to train the model, the examples from the end of 2015 to the end of 2016 are used for validation, and the examples from the end of 2016 to the end of 2017 are used to test and assess the model. The split of training data into contiguous pieces is of paramount importance for time series in general, and for this problem in particular. For instance, applying a random train/test split would assign pieces of any tremor burst longer than a few minutes to both the training and the testing set, making the problem dramatically easier -- the CNN would only have to memorize the examples it sees in training. 

The architecture of the CNN consists of three convolutional layers and one fully connected hidden layer (see Fig 1 and Methods for details). At the end of the training procedure (Fig. S1) the model is fixed and applied to the testing set (the last year of data never analyzed before), to assess how well it generalizes to new examples of tremor. \\

The CNN outputs an empirical probability that a portion of waveform transformed into the time-frequency domain contains tremor. The performance of the model, measured through the ROC-AUC metric, is shown in Fig. 2. The ROC curve is generated by plotting true positive rate of the model on the y-axis versus the false positive rate on the x-axis at progressively larger threshold settings. The further the curve is from the diagonal line and closer to the upper left corner, the better the model is at discriminating between positives and negatives. Lowering the classification threshold classifies more items as positive, thus increasing both false positives and true positives. For our purposes, true positives are catalogued events %(identified from multi-station envelope correlation, never seen in training by our model) 
detected on a single station by our deep learning model. False positives are detections from our model of possible events that were not catalogued. Figure 2B shows the model performance on the test set (last year of the labeled database, 2017) according to the ROC curve: with a ROC-AUC score of 0.945, the model performs well in discriminating between positives (catalogued events) and negatives (waveforms not catalogued, and on days with no catalogued events). The confusion matrix in inset shows the fraction of classified noise and tremor compared to the actual labels, for a model with a default threshold of 0.5. \\

\subsection*{Tremorness}

\begin{figure}[!ht]
\begin{center}
\includegraphics[width=14cm,trim= 0 0 0 0]{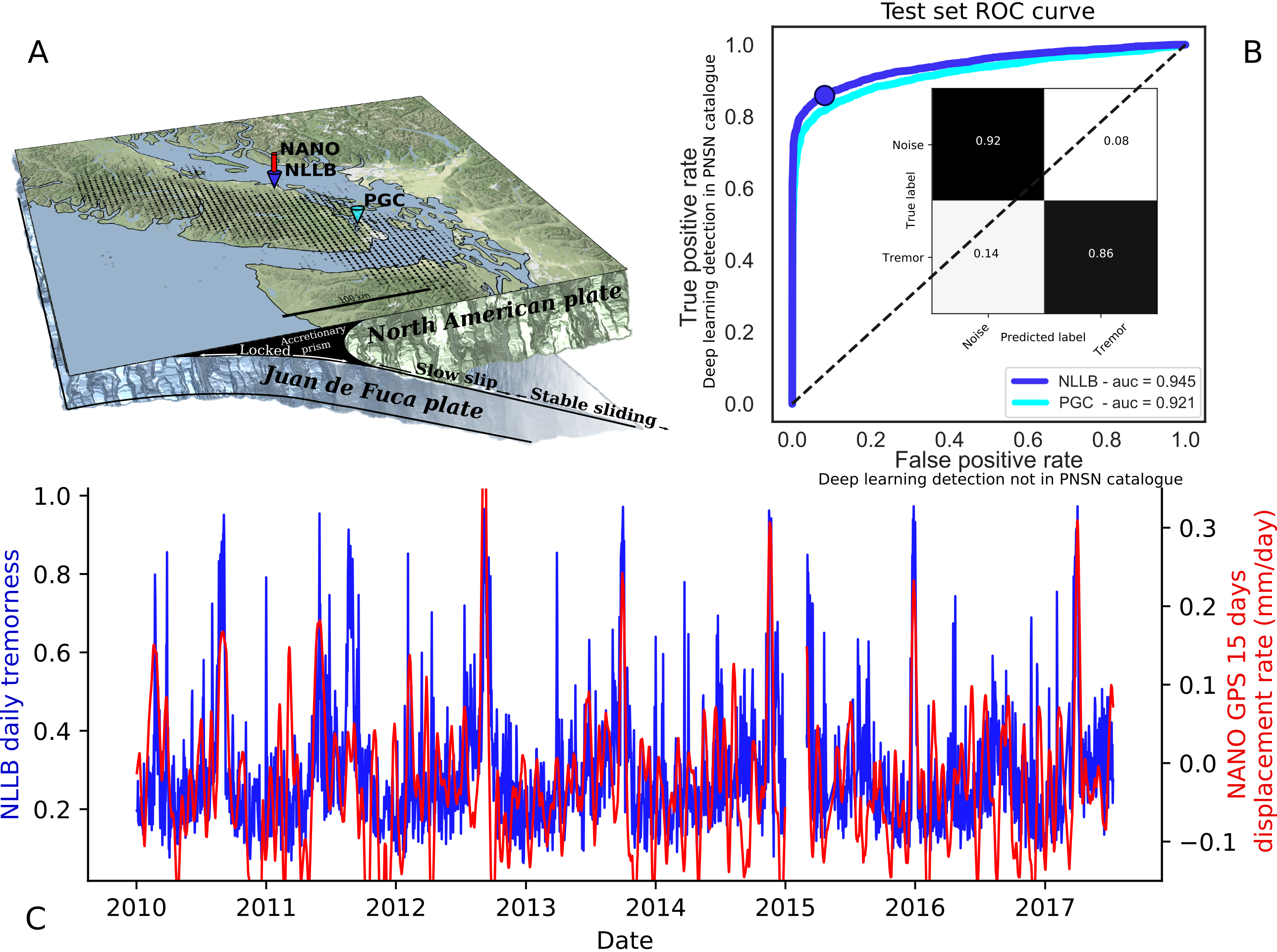}
\caption{\footnotesize{\textbf{`Tremorness' continuously tracks geodetic slow slip displacement rate, and generalizes to nearby stations. } \textbf{(A)} Map of Vancouver Island. Slow slip and tremor originate from ductile portions of the interface, down dip from the locked zone where a megathrust earthquake is anticipated. Seismic stations NLLB and PGC as well as the GPS station NANO are noted by colored arrows.  \textbf{(B)}  The area under the receiving operating characteristic (ROC) curve and the confusion matrix. The ROC curve shows the true (deep learning detection of catalogued event) and false (deep learning detection of a possible but uncatalogued event) positive rates as the threshold of classification of the model is varied. A model that reproduces the catologue exactly would yield a point in the upper left corner or coordinate (0,1) of the ROC space.   The inset shows the confusion matrix, indicating the fraction of classified noise and tremor by the deep learning model compared to the labels from the multi-station catalogue, for a model with a threshold of 0.5. Most tremor catalogued using multi-station cross-correlations are identified on a single station by the neural network. Other signals such as earthquakes, teleseisms, cultural noise and microseisms are easily distinguished from tremor by the model (see also Fig. S2). \textbf{(C)} Blue: daily average of the tremor content of the seismic data from the NLLB station, as determined by the deep learning model. Red: 15 days average of the co-located GPS displacement rate, from station NANO. }}
\label{fig2}
\end{center}
\end{figure}

Given continuous seismic spectral data as input, the deep learning model outputs an empirical probability that the seismic data contains tectonic tremor. We term this empirical probability `tremorness'. We showed in previous work that the energy of continuously-recorded low-amplitude seismic waves track the smoothed GPS displacement rate well at long time scales (30+ days) and short time scales (one hour) \cite{Rouet2019}. The energy-GPS correlation may suggest that tremor is emitted continuously or quasi-continuously at the plate interface from its slowly slipping portion (Fig. 2A). In contrast, catalogued tremor has been shown to be intermittent \cite{WechBartlow}. Using the deep CNN trained to recognize catalogued tremor, we find strong evidence that tremor is emitted at least quasi-continuously, and is a quantitative proxy for geodetic slow slip.

Figure 2C shows that the daily tremorness characteristic of the seismic data from the NLLB station tracks the co-located SW GPS displacement rate, even at small displacement rates and with modest GPS smoothing. In between the peaks of slow slip rate, the deep learning model finds that the seismic data on any given day contains 15 to 40\% tremor (compared with 0.08\% where no slow slip occurs, the actual false positive rate of our model, see Fig. S2). These weak tremors that go undetected in the multi-station method map to smaller peaks in the GPS displacement rate (see Fig. S4), demonstrating the possible existence of numerous slip events in between the known large slow slip events, as have been observed for instance in Mexico \cite{Frank2016}. %This result suggests that the continuous tremor-like signal identified in our previous work \cite{Rouet2019} is in fact classical tremor.
If we consider a detection threshold of tremor for a tremorness (CNN output) above 0.5 empirical probability, the default classifier built by the deep learning model, we detect more than 130,000 5-minute waveforms containing tremor events, close to three times the number contained in the catalog for the same time interval. 

We note that our neural network has been trained to reproduce the catalogue as well as it could using a single seismic station. This corresponds to a threshold of 0.5 on the ROC curve on Fig. 2B. The deep learning model of tremor is therefore conservatively trained and the additional events it detects do not come from lowering the standards of detection compared to the multi-station method. This means that the model finds more events than those in the catalogue. This finding is not because these events correspond to a lower detection standard, but because according to the spectral features learned on a single station they are as likely to be tremor as the known catalogued events. These events are too weak in amplitude (see Fig. S3) to be detected on multiple stations, but based on the temporal evolution of their frequency content, the deep learning model cannot distinguish these newly detected tremor events from the known catalogued events. The correlation of tremorness and GPS displacement rate (Fig. 2C and S4) further demonstrates that these detections are not spurious. As a further test we applied the trained model to seismic data measured in a region known to be seismically inactive, in this case Michigan, USA (Fig., S2). Extremely little tremor is identified in Michigan by our model, supporting that the numerous newly detected tremor events in Cascadia are real. It also indicates that our model is not confused by cultural noise (e.g., train traffic, vehicular traffic, wind farms), meterological noise (wind, storms), or teleseisms. \\

\section*{Discussion}

\subsection*{Deep learning model interpretation}

Deep learning models are notoriously hard to interpret. However, recent efforts \cite{Baehrens2010,Montavon2017} have showed that perturbing the input of deep learning models enables their analysis, to some extent. In our case the network outputs the empirical probability that the input contains tremor, and a Taylor expansion of the model reveals the time-frequency components the network used to classify the signal as tremor.
Fig.1D shows examples of such an analysis, applying a Taylor expansion of the neural network with respect to its input pixels \cite{Montavon2017} to construct a heatmap of time-frequency components identified used by the deep learning model to identify tremor. Fig.1E shows the inverse short-time Fourier transform of the time-frequency components identified as tremor by the Taylor analysis. Here, because the Taylor expansion identifies individual time-frequency components used to identify tremor, we can reconstruct a signal that represents the separation of the tremor signal from the background noise. We caution that this signal extraction method is only partial, as the Taylor expansion only reveals the time-frequency components that are characteristic of tremor, and not components that may be common between tremor and other signals. \\

% Here, because each pixel of our input has a physical meaning as a time-frequency component of the seismic signal, activation of the CNN has a direct physical interpretation and the tremor signal can be separated from the background seismic data. %This approach offers a powerful means to identify tremor temporal and frequency characteristics. 
% Fig.1E shows the tremor signals identified by the deep learning model, reconstructed by inverse short-time Fourier transform the time-frequency components of the initial STFT identified as tremor by the Taylor analysis.

The features learned by the convolutional neural network are patterns in the time-frequency domain, and we posit these features are directly related to the frictional properties of the slip on the interface that emit the signals. We presume that variations in pore pressure, chemistry, or thermal properties may modulate or influence the emitted signal, but the origin of the signal is due to emissions coming from asperities on the fault interface. In Fig. S5 we show  evidence of the link between features learned by the CNN and inferred variations in frictional properties of the fault. The last feature map of the CNN, computed for classically catalogued and located tremor, exhibits a clustering in feature space that is reflected in its geographic location. This relation suggests that the tremor features learned by the CNN may be specific to the frictional properties of slip, as these features evolve systematically with the spatial origin of tremor. It is also possible the clustering results from distinct differences in path effects. This is a point we intend to explore further in future work, along with evolution of these features with time \cite{Holtzman2018}. \\
 
\subsection*{Universality of tremor characteristics}

\begin{figure}[!ht]
\begin{center}
\includegraphics[width=12cm,trim= 0 0 0 0]{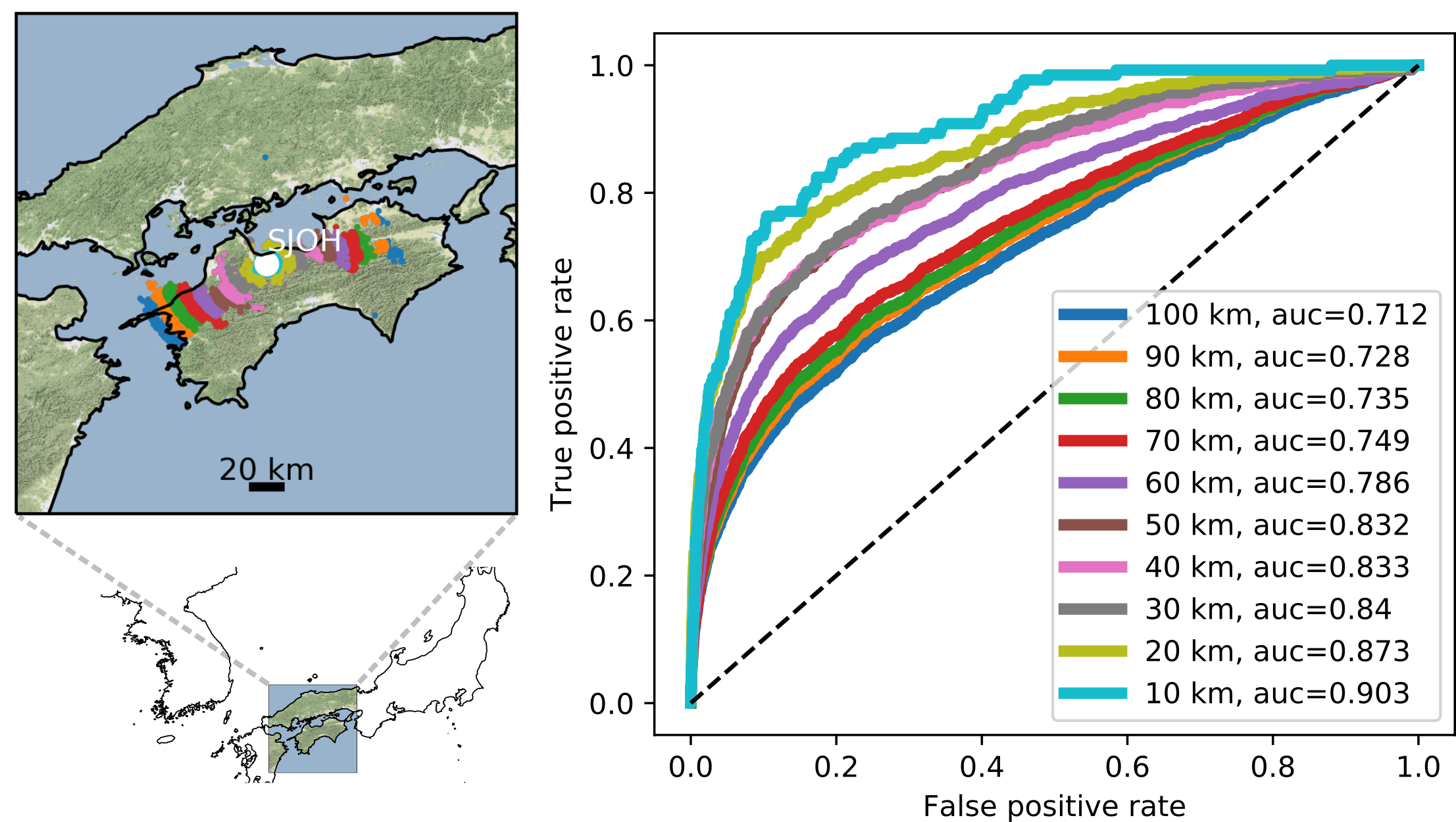}\caption{\footnotesize{\textbf{The deep neural network model of tremor trained in Cascadia recognizes known tremor in other regions.} Left: Maps of catalogued tremor in Japan (A) and California (B). Right: ROC curves showing the accuracy of our model trained in Cascadia at recognizing tremor in Japan (A) and California (B), with colors matching. \textbf{(A)} ROC curves of the deep learning model of tremor exclusively trained in Cascadia, and applied to tremor catalogued in Shikoku, Japan \cite{Ide2012,Idehara2014}. The performance decreases with distance (colors on maps and ROC curves matching). At short distances ($<$20 km) the model has the same performance on Japanese tremor as it does on Cascadia tremor. \textbf{(B)} ROC curves of the deep learning model applied to tremor catalogued on the San Andreas fault \cite{Shelly2017}. The deep model trained on the Cascadia subduction zone only could not be used to build catalogues there on the San Andreas strike slip fault, but its ability to recognize catalogued tremor in most cases underscores the frictional similarity between strike slip tremor and subduction tremor. }}
\label{fig3}
\end{center}
\end{figure}

Deep learning models have tremendous expression capabilities, which can lead to overfitting. In the previous section we demonstrated that our network generalizes to new seismic data from the station it was trained on, recorded later in time, suggesting that overfitting is not an issue. A powerful additional test is to see whether the network generalizes to seismic data from another station. Seismic recordings at a given station are a convolution of the source, propagation path and `site amplification' effects, that may vary tremendously over short distances \cite{aki2002quantitative}. Generalization to another station would suggest that the network did not only learn the specifics of the seismic data it has been trained on. In Fig. 2B we show that our trained network is robust when applied to other seismic stations on Vancouver Island. The network trained on station NLLB from the end of 2009 to the end of 2015 can accurately recognize catalogued tremor on station PGC from 2016 to 2017, 80 km away. This test supports that the network did learn general time-frequency dynamics that are characteristic of tremor.

Furthermore, the model generalizes to other subduction zones and different tectonic environments (Fig. 3). %, providing strong evidence for universal frictional characteristics driving slow slip and tremor. In Fig. 3 we show that tremor signals are not only specific and distinctive, but generalizes to other regions. 
Fig. 3A shows that catalogued tremor in Shikoku, southern Japan\cite{Ide2012,Idehara2014} (also developed using multi-station envelope correlation) is accurately identified on a single station by our deep learning model of tremor trained in Cascadia, with no further training required. Fig. 3B shows that known tremor emitted by the San Andreas transform fault \cite{Shelly2017} -- a very different tectonic environment -- is also identified by the same deep learning model trained in Cascadia. Tremor on the San Andreas fault has recently been shown to be due to slow slip on the deep portion of the fault \cite{Rousseteaav3274}, similar to slow slip in subduction zones. These results show that the time-frequency dynamics of tremor signals are largely dominated by the source characteristics, because our model is station and even region agnostic. \\

If tremor signals are dictated by source characteristics, then the time-frequency dynamics of these signals are indicative of an event rate and distribution that is characteristic of tremor: bursts of discrete events that are very similar in magnitude (very high b-values), with a very high event rate given by the time-frequency evolution of detected tremor. From this perspective, the time-frequency dynamics of tremor are a fingerprint of the frictional properties of the asperities that emit them.
 The ability of our model to recognize tremor in Japan and from the San Andreas fault suggests that the waveform characteristics of tremor are universal.  This result is in line with laboratory analysis \cite{Scuderi2016} and models \cite{Daub2011} -- tremor-inducing slow slip occurs within frictional conditions that are very similar for a wide variety of faults, and possibly all faults systems.

\section*{Conclusion}

A deep learning model can be trained in Cascadia to accurately detect tectonic tremor on a single station.
Trained on catalogued tremor events initially identified by multi-station methods, the deep learning model can be used to find many more events. The neural network gives a continuous measure of tremor content, that tracks geodetic displacement in Cascadia and provides clearer time bounds on slow slip events. Deep learning detection of tremor rises months before slow slip is detected geodetically.
Trained only in Cascadia, the deep learning model recognizes known tremor from other subduction zones as well as from the deep portion of the San Andreas fault, arguing for the universality of the frictional properties governing slow earthquakes.

\acknowledgments
This work was funded by Institutional Support (LDRD) at Los Alamos and the DOE Office of Basic Research, Geoscience Program. We thank Romain Jolivet, Joan Gomberg and Daniel Trugman for helpful discussions. We thank Tim Cote, Xiuying Jin and Michal Kolaj from the Canadian National Seismograph Network for helping us obtain the data and for data troubleshooting. The seismic data used were obtained from the Canadian National Seismograph Network \cite{CNSN}, and the GPS data are from the Western Canada Deformation Array (WCDA), processed by the USGS Pacific Northwest NetworkS (doi:10.5066/F7NG4NRK). This work was supported by DOE Office of Science (Geoscience Program) grant 89233218CNA000001. CH was partly financed by CEA/DASE (and hosted at the Yves Rocard Joint Laboratory ENS‐CNRS‐CEA/DASE), and by the European Research Council (ERC) under the European Union's Horizon 2020 research and innovation program (Geo-4D project, grant agreement 758210).

\textbf{Author Contributions}. BRL devised the original study and conducted the machine learning work. BRL, CH, IM and PAJ analyzed and interpreted the results. \\

\textbf{Financial and non-financial competing interests}. The authors declare no competing financial or non-financial interests.

\section*{Data and code availability} The data used is publicly available and can be found online. The seismic data come from the Canadian National Seismograph Network \cite{CNSN} (www.earthquakescanada. nrcan.gc.ca), and the GPS data come from the Western Canada Deformation Array (WCDA) operated by the Geological Survey of Canada (GSC), preprocessed by the USGS \cite{USGS} \\(https://earthquake.usgs.gov/monitoring/gps/Pacific\_Northwest, NA-fixed trended data). The known tremor catalog used is available from the Pacific Northwest Seismic Network (https://pnsn.org/tremor/tremor-map-legacy).

The work flow described in the Methods uses open source software: python and python packages including scikit-learn \cite{Pedregosa2011}, tensorflow (https://www.tensorflow.org/), keras (https://keras.io/)  and obspy \cite{obspy}.

\end{document}